\documentclass[prd,a4paper,showpacs,preprint,byrevtex]{revtex4}
\usepackage{dcolumn}
\usepackage{amsmath}
\usepackage{bm}

\begin{document}

\title{Auxiliary fields and the flux tube model}

\author{Fabien \surname{Buisseret}}
\author{Claude \surname{Semay}}
\thanks{FNRS Research Associate}
\email[E-mail: ]{claude.semay@umh.ac.be}
\affiliation{Service de Physique G\'{e}n\'{e}rale et de Physique des
Particules \'{E}l\'{e}mentaires, Groupe de Physique Nucl\'{e}aire
Th\'{e}orique,
Universit\'{e} de Mons-Hainaut, Place du Parc 20,
B-7000 Mons, Belgium}

\date{\today}

\begin{abstract}
It is possible to eliminate exactly all the auxiliary fields (einbein
fields) appearing in the rotating string Hamiltonian to obtain the
classical equations of motion of the relativistic flux tube model. A
clear interpretation can then be done for the characteristic variables
of the rotating string model.
\end{abstract}


\maketitle

It has been recently shown that the solutions of the Nambu-Goto
Lagrangian for two quarks attached with a straight string are given by
the solutions of the relativistic flux tube model \cite{alle03}. To get
rid of the square roots appearing in this Lagrangian, a Hamiltonian for
a rotating string containing auxiliary fields (also known as einbein
fields) was previously written from this Lagrangian
\cite{morg99,guba94}. It has been shown that, in the case of two equal
quark masses, the equations of motion of the rotating string reduce
exactly to the classical equations of motion of the relativistic
flux tube model, provided all auxiliary fields are eliminated correctly
\cite{sema04}. In this work, we generalize this result to the case of
two quarks with different masses, and we provide a clear interpretation
for the characteristic variables of the rotating string model.

Partial results about the elimination of the auxiliary fields for the
rotating string Hamiltonian have been obtained in Ref.~\cite{guba94},
but we perform here the complete calculation. Moreover, we
show that some approximations which seems to be used in
Ref.~\cite{guba94} are not necessary.

The Lagrangian for a system of two (spinless) quarks with masses $m_1$
and $m_2$ attached with a string can be written \cite{morg99}
\begin{equation}
\label{nambu}
{\cal L}(\tau)=-m_{1}\sqrt{\dot{x}^{2}_{1}}-m_{2}\sqrt{\dot{x}^{2}_{2}}-
a\int^{1}_{0}d\beta\sqrt{(\dot{w}w')^{2}-\dot{w}^{2}{w'}^{2}},
\end{equation}
where $a$ is the string tension. $x_1(\tau)$ and $x_2(\tau)$ are the
world lines of the two quarks, which depend on the common proper time
$\tau_1=\tau_2=\tau$. The four-vector $w(\tau,\beta)$ is the string
coordinate,
parameterized by $\beta\in[0,1]$. The notations $\dot{x}=dx/d\tau$ and
$w'=dw/d\beta$ are used. The introduction of auxiliary fields to get rid
of the square roots leads to the following Lagrangian
\begin{eqnarray}
\label{step1}
{\cal L}&=&-\frac{1}{2}\left\{\frac{m^{2}_{1}}{\mu_{1}}+\mu_{1}
\dot{x}^{2}_{1}+
\frac{m^{2}_{2}}{\mu_{2}}+\mu_{2}\dot{x}^{2}_{2}\right.
\nonumber \\
&&+\left.\int^{1}_{0} d\beta\  \nu \left[\dot{w}^{2}-\frac{a^{2}}
{\nu^{2}}{w'}^{2}-2\eta(\dot{w}w')+\eta^{2}{w'}^{2}\right]\right\}.
\end{eqnarray}
As we will see later, the field $\mu_i(\tau)$ can be considered as a
constituent mass for the quark $i$, while the field $\nu(\tau,\beta)$
plays the role of a string energy density. The field $\eta(\tau,\beta)$
is the easiest field to eliminate. In the following, we will
adopt the straight-line ansatz for the minimal string
\begin{equation}
\label{recti}
w(\tau,\beta)=\beta x_1(\tau)+(1-\beta) x_2(\tau).
\end{equation}

We define a set of relative and center of mass coordinates
\begin{subequations}
\label{cmcoord}
\begin{eqnarray}
\label{cmr}
r(\tau)&=&x_1(\tau)-x_2(\tau), \\
\label{cmgr}
R(\tau)&=&\zeta(\tau) x_1(\tau)+(1-\zeta(\tau)) x_2(\tau),
\end{eqnarray}
\end{subequations}
The parameter $\zeta(\tau)$ will be determined below. With these
new coordinates, the condition~(\ref{recti}) is written (we will now
drop the notations of the various dependencies of the fields on $\tau$
and $\beta$)
\begin{equation}
\label{wdef}
w=R+(\beta-\zeta) r,
\end{equation}
and the Lagrangian~(\ref{step1}) becomes
\begin{eqnarray}
\label{step2}
{\cal L}&=& -\frac{1}{2} \left[
\frac{m^{2}_{1}}{\mu_{1}}+\frac{m^{2}_{2}}{\mu_{2}}+
a_{1}
\dot{R}^{2}+2a_{2}\dot{R}\dot{r}-2(c_{1}+\dot{\zeta}a_{1})\dot{R}r-2(
c_{2}+\dot{\zeta}a_{2})\dot{r}r \right.\nonumber \\
&&\left.
\phantom{-\frac{1}{2}}+a_{3}\dot{r}^{2}+(a_{4}+2\dot{\zeta}c_{1}+\dot{
\zeta}^{2}a_{1})r^{2}
\right].
\end{eqnarray}
The following notations are used
\begin{subequations}
\label{acdef}
\begin{eqnarray}
a_{1}&=&\mu_{1}+\mu_{2}+\int^{1}_{0} d\beta\, \nu , \label{a1def} \\
a_{2}&=&\mu_{1}-\zeta(\mu_{1}+\mu_{2})+\int^{1}_{0} d\beta \,
(\beta-\zeta)\, \nu , \label{a2def} \\
a_{3}&=&\mu_{1}(1-\zeta)^{2}+\mu_{2}\zeta^{2}+\int^{1}_{0} d\beta\,
(\beta-\zeta)^{2}\, \nu, \label{a3def} \\
a_{4}&=&\int^{1}_{0} d\beta\,
\left(\eta^{2}\nu-\frac{a^{2}}{\nu}\right),
\label{a4def} \\
c_{1}&=&\int^{1}_{0} d\beta \, \eta\,  \nu, \label{c1def} \\
c_{2}&=&\int^{1}_{0} d\beta\,  (\beta-\zeta)\,  \eta\,  \nu.
\label{c2def}
\end{eqnarray}
\end{subequations}
We can extract three useful relations
\begin{subequations}
\label{intuse}
\begin{eqnarray}
\label{def1}
\int^{1}_{0} d\beta\, \nu&=&a_{1}-\mu_{1}-\mu_{2} , \\
\label{def2}
\int^{1}_{0} d\beta \, \beta\, \nu&=&a_{2}+\zeta a_{1}-\mu_{1}, \\
\label{def3}
\int^{1}_{0} d\beta\, \beta^{2}\, \nu&=&a_{3}+\zeta^{2}a_{1}+
2\zeta a_{2}-\mu_{1}.
\end{eqnarray}
\end{subequations}

In the following, the contravariant components of a four-vector
$v$ will be noted $(v^0,\vec v\,)$ and $v^2=(v^0)^2-\vec v\,^2$.
The use of a common proper time implies that $r^0=\dot{r}^0=0$ and that
$\dot{R}^0=1$. Consequently, the Lagrangian can be recast into
\begin{eqnarray}
\label{step3}
{\cal L}&=& -\frac{1}{2} \left[
\frac{m^{2}_{1}}{\mu_{1}}+\frac{m^{2}_{2}}{\mu_{2}}+a_{1}-
a_{1}\dot{\vec{R}}^{\, 2}-2a_{2}\dot{\vec{R}}\dot{\vec{r}}+2(c_{1}+\dot{
\zeta}a_{1})\dot{\vec{R}}\vec{r}+2(c_{2}+\dot{\zeta}a_{2})\dot{\vec{r}}
\vec{r} \right. \nonumber \\
&&\phantom{-\frac{1}{2}}\left. -a_{3}\vec{r}^{\, 2}-(a_{4}+2\dot{\zeta}
c_{1}+\dot{\zeta}^{2}a_
{1}) \vec{r}^{\, 2}\right].
\end{eqnarray}

As we will work in the center of mass frame, the total vector momentum
$\vec{P}_{\text{cm}}$ of the system must vanish
\begin{equation}
\label{pcm}
\vec{P}_{\text{cm}}=-\frac{\partial{\cal
L}}{\partial\dot{\vec{R}}}=\vec{0},
\end{equation}
which implies that
\begin{equation}
\label{cm_rep}
\dot{\vec{R}}=\frac{(c_{1}+\dot{\zeta}a_{1})\vec{r}-a_{2}\dot{\vec{r}}}
{a_{1}}.
\end{equation}
Moreover, the relative vector momentum $\vec{p}$ is given by
\begin{equation}
\label{p_rel}
\vec{p}=-\frac{\partial{\cal
L}}{\partial\dot{\vec{r}}}=-a_{2}\dot{\vec{R}}+(c_{2}+\dot{\zeta}a_{2})
\vec{r}-a_{3}\dot{\vec{r}}.
\end{equation}
Thus, we impose $a_2=0$ in order that $\vec{p}$ does not depend on the
motion of the center of mass. This leads to the following value for
the parameter $\zeta$
\begin{equation}
\label{zeta}
\zeta=\frac{\mu_{1}+\int^{1}_{0} d\beta\, \beta\,
\nu}{\mu_{1}+\mu_{2}+\int^{1}_{0} d\beta\, \nu}.
\end{equation}
This quantity, which determines the center of mass coordinate, contains
contributions not only from the quarks but also from the string.
From this expression, it is clear that fields $\mu_1$ and $\mu_2$ play
the role
of dynamical masses for the quarks, while the field $\nu$ can be
interpreted as a dynamical string energy density.
The substitution of $\dot{\vec{R}}$ by the formula~(\ref{cm_rep}) and
the replacement of $a_2$ by zero give the following Lagrangian
\begin{equation}
\label{lagr_1}
{\cal L}= -\frac{1}{2} \left[
\frac{m^{2}_{1}}{\mu_{1}}+\frac{m^{2}_{2}}{\mu_{2}}+a_{1}+
\frac{1}{a_{1}}\left((c^{2}_{1}-a_{4}a_{1})\vec{r}^{\, 2}-a_{3}a_{1}\dot
{\vec{r}}^{\, 2}+2a_{1}c_{2}\dot{\vec{r}}\vec{r}\right) \right].
\end{equation}
It is worth noting that all terms with $\dot{\zeta}$ have disappeared.
It
is thus not necessary to assume $\zeta$ independent of $\tau$ as it
seems to be the case in Ref.~\cite{guba94}.

The Lagrangian~(\ref{lagr_1}) contains auxiliary fields through the
variables $a_i$ and $c_i$. The field $\eta$ can be first eliminated by
the condition $\delta{\cal L}/\delta\eta=0$ \cite{guba94}. This gives
the following expression
\begin{equation}
\label{var_eta}
2(\dot{\vec{r}}\vec{r}\,)\delta c_{2}+\frac{\vec r\,^2}
{a_{1}}\left[2c_{1}\delta c_{1}-a_{1}\delta a_{4}\right]
=0.
\end{equation}
With relations~(\ref{intuse}), this equation can be recast into the form
\begin{equation}
\label{eta}
c_{1}=\frac{\dot{\vec{r}}\vec{r}}{
\vec{r}^{\, 2}}a_{1}\left(\zeta-\frac{\mu_{1}}{\mu_{1}+\mu_{2}}\right).
\end{equation}
The use of Eq.~(\ref{a1def}), with the condition $a_2=0$, gives the
solution
\begin{equation}
\label{eta0}
\eta_{0}=\frac{\dot{\vec{r}}\vec{r}}{\vec{r}^{\, 2}}
\left(\beta-\frac{\mu_{1}}{\mu_{1}+\mu_{2}}\right).
\end{equation}
Knowing the extremal value of the auxiliary field $\eta_0$, the
coefficients $c_1$, $c_2$, and $a_4$ can be computed. We thus obtain
\begin{subequations}
\label{c2c1a4a1}
\begin{eqnarray}
c_{2}&=&\frac{\dot{\vec{r}}\vec{r}}{\vec{r}^{\,
2}}(a_{3}-\mu), \label{c2new} \\
c^{2}_{1}-a_{4}a_{1}&=&a_{1}\left[\int^{1}_{0}d\beta \frac{a^{2}}{\nu}-
(a_{3}-\mu)\frac{(\dot{\vec{r}}\vec{r}\,)^{2}}{\vec{r}^{\,
4}}\right], \label{c2a4a1new}
\end{eqnarray}
\end{subequations}
where $\mu=\mu_{1}\mu_{2}/(\mu_{1}+\mu_{2})$.

It is now interesting to separate the longitudinal and transverse
components of $\dot{\vec r}$ with respect to $\vec r$
\begin{equation}
\label{rparperp}
\dot{\vec{r}}^{\, 2}=
\dot{\vec{r}}\,^{2}_{\parallel}+\dot{\vec{r}}\, ^{2}_{\bot} =\frac{(
\dot{\vec{r}}\vec{r}\, )^{2}}{\vec{r}^{\,
2}}+\frac{(\dot{\vec{r}}\times\vec{r}\, )^{2}}{\vec{r}^{\, 2}},
\end{equation}
to obtain
\begin{equation}
\label{lagrparperp}
{\cal L}=-\frac{1}{2}\left[\frac{m^{2}_{1}}{\mu_{1}}+\frac{m^{2}_{2}}{
\mu_{2}}+a_{1}+a^{2}\vec{r}^{\, 2}\int^{1}_{0} \frac{d\beta}{\nu}-a_{3}
\dot{\vec{r}}\, ^{2}_{\bot}-\mu\dot{\vec{r}}\, ^{2}_{\parallel}
\right].
\end{equation}
The canonical transformation
$H=-\dot{\vec{r}}_{\parallel}\vec{p}_{r}-\dot{\vec{r}}_{\bot}\vec{p}
_{\bot}-{\cal L}$, with
\begin{subequations}
\label{canon}
\begin{eqnarray}
\vec{p}_{r}&=&-\frac{\partial{\cal L}}{\partial
\dot{\vec{r}}_{\parallel}}=-
\mu\dot{\vec{r}}_{\parallel}, \label{pr} \\
\vec{p}_{\bot}&=&-\frac{\partial{\cal L}}{\partial
\dot{\vec{r}}_{\bot}}=-
a_{3}\dot{\vec{r}}_{\bot}, \label{pperp}
\end{eqnarray}
\end{subequations}
lead to the following Hamiltonian (now, $r^2$ always means $\vec r\,
^2$)
\begin{equation}
\label{QCD_eq1}
H=\frac{1}{2}\left[\frac{p^{2}_{r}+m^{2}_{1}}{\mu_{1}}+\frac{p^{2}_{r}+m
^{2}_{2}}{\mu_{2}}+\mu_{1}+\mu_{2}+a^{2}r^{2}\int^{1}_{0}\frac{d\beta}{
\nu}+
\int^{1}_{0}d\beta\nu+\frac{L^{2}}{a_{3}r^{2} } \right],
\end{equation}
where the orbital angular momentum is given by the usual relation
$L^{2} = \vec{p}\,^2_{\bot} r^2$. Let us note that the
string does not contribute to the longitudinal component of the momentum
\cite{guba94}.

We will now proceed to the elimination of the auxiliary field $\nu$ by
demanding that $\delta H/\delta\nu=0$ \cite{guba94}. This leads to the
following expression
\begin{equation}
\label{deltahnu}
\int^{1}_{0}d\beta\left(1-\frac{a^{2}r^{2}}{\nu^{2}}\right)\delta\nu
-\frac{L^{2}}{a^{2}_{3}r^{2}}\delta a_{3}=0.
\end{equation}
Using the definition~(\ref{a3def}) and the condition $a_2=0$
(\ref{zeta}), we obtain
\begin{equation}
\label{a3cond}
\delta a_{3}=-2a_{2}\delta \zeta+
\int^{1}_{0}d\beta(\beta-\zeta)^{2}\delta \nu=\int^{1}_{0}d\beta(\beta-
\zeta)^{2}\delta \nu.
\end{equation}
So, we have to solve
\begin{equation}
\label{numin}
\int^{1}_{0}d\beta\ \delta\nu\left(1-\frac{a^{2}r^{2}}{\nu^{2}}-\frac{L^
{2}}{a^{2}_{3}r^{2}}(\beta-\zeta)^{2}\right)=0.
\end{equation}
In order to simplify the notation, we introduce the variable $y$
\begin{equation}
\label{ydef}
y^{2}=\frac{L^{2}}{a^{2}_{3}r^{2}}.
\end{equation}
The extremal field $\nu_0$ solution of Eq.~(\ref{numin}) is
\begin{equation}
\label{nu0}
\nu_{0}=\frac{ar}{\sqrt{1-y^{2}(\beta-\zeta)^{2}}}.
\end{equation}
It is now possible to compute the various coefficients present in the
Hamiltonian~(\ref{QCD_eq1}).
The following integrals are given by analytical formulas
\begin{eqnarray}
\label{rel3}
\int^{1}_{0}d\beta \nu&=&\frac{ar}{y}\left[\arcsin
s\right]^{y_{1}}_{-y_{2}}, \\
\label{rel4}
\int^{1}_{0}\frac{d\beta}{\nu} &=&\frac{1}{2ary}\left[s\sqrt{1-s^{2}}+
\arcsin s\right]^{y_{1}}_{-y_{2}}, \\
\label{rel1}
\int^{1}_{0}d\beta
(\beta-\zeta)\nu&=&-\frac{ar}{y^{2}}\left[\sqrt{1-s^{2}}\right]^{y_{1}}_
{-y_{2}},\\
\label{rel2}
\int^{1}_{0}d\beta
(\beta-\zeta)^{2}\nu&=&\frac{ar}{2y^{3}}\left[-s\sqrt{1-s^{2}}+\arcsin s
\right]^{y_{1}}_{-y_{2}},
\end{eqnarray}
with the notations
\begin{equation}
\label{y1y2}
y_{1}=(1-\zeta) y \quad \text{and} \quad y_{2}=\zeta y.
\end{equation}

Using the constraint $a_2 y = 0$ (\ref{zeta}), the relation
$a_3 y = L/r$ (\ref{ydef}), and the Hamiltonian (\ref{QCD_eq1}), we
obtain a set of three coupled equations for the rotating string
\begin{subequations}
\label{seteq1}
\begin{eqnarray}
0&=&\mu_{1}y_{1}-\mu_{2}y_{2}-\frac{ar}{y}\left(\sqrt{1-y^{2}_{1}}-\sqrt
{1-y^{2}_{2}}\right), \label{peq0} \\
\frac{L}{r}&=&\frac{1}{y}(\mu_{1}y^{2}_{1}+\mu_{2}y^{2}_{2})+\frac{ar}
{y^{2}} (F(y_{1})+F(y_{2})), \label{lor} \\
H&=&\frac{1}{2}\left[\frac{p^{2}_{r}+m^{2}_{1}}{\mu_{1}}+\frac{p^{2}_{r}
+m^{2}_{2}}{\mu_{2}}+\mu_{1}(1+y^{2}_{1})+\mu_{2}(1+y^{2}_{2})\right]
\nonumber \\
&&+\frac{ar}{y}(\arcsin y_{1}+\arcsin y_{2}), \label{haux}
\end{eqnarray}
with
\begin{equation}
\label{fy}
F(y_{i})=\frac{1}{2}\left[\arcsin y_{i}-y_{i}\sqrt{1-y^{2}_{i}}
\right].
\end{equation}
\end{subequations}

The elimination of the remaining auxiliary fields $\mu_1$ and $\mu_2$ is
obtained using the constraints $\delta H/\delta\mu_{i}=0$
\cite{guba94}. As in Ref.~\cite{sema04}, we also ask
$\delta (L/r)/ \delta\mu_{i}=0$. So we can write
\begin{equation}
\label{cond0}
2\partial_{\mu_{i}}H-2y\partial_{\mu_{i}} (L/r)=0.
\end{equation}
These equations lead to the following relations
\begin{subequations}
\label{eq0102}
\begin{eqnarray}
-\frac{p^{2}_{r}+m^{2}_{1}}{\mu^{2}_{1}}+1-y^{2}_{1}+2A\partial_{\mu_{1
}}y_{1}+2B   \partial_{\mu_{1}}y_{2}&=&0, \label{eq01} \\
-\frac{p^{2}_{r}+m^{2}_{2}}{\mu^{2}_{2}}+1-y^{2}_{2}+2B\partial_{\mu_{2
}}y_{1}+2A   \partial_{\mu_{2}}y_{2}&=&0, \label{eq02}
\end{eqnarray}
\end{subequations}
with
\begin{subequations}
\label{gagb}
\begin{eqnarray}
A&=&-\mu_{1}y_{1}+\frac{ar}{y}\sqrt{1-y^{2}_{1}}+\frac{1}{y}(\mu_{1}y^{2
}_{1} +\mu_{2}y^{2}_{2})-\frac{ar}{y^{2}}\left(y_{1}\sqrt{1-y^{2}_{1}}+
y_{2}\sqrt{1-y^{2}_{2}}\right),\label{ga} \\
B&=&-\mu_{2}y_{2}+\frac{ar}{y}\sqrt{1-y^{2}_{2}}+\frac{1}{y}(\mu_{1}y^{2
}_{1} +\mu_{2}y^{2}_{2})-\frac{ar}{y^{2}}\left(y_{1}\sqrt{1-y^{2}_{1}}+
y_{2}\sqrt{1-y^{2}_{2}}\right).\label{gb}
\end{eqnarray}
\end{subequations}
It is easy to see that $B-A=a_2 y=0$. Moreover, using
definitions~(\ref{y1y2}), it can be shown that $A=-\zeta a_2 y=0$.
Finally, $A=B=0$
and the extremal values of the auxiliary fields $\mu_i$ are given by
\begin{equation}
\label{mui0}
\mu_{i0}=\sqrt{\frac{p^{2}_{r}+m^{2}_{i}}{1-y^{2}_{i}}}.
\end{equation}
If we interpret the variables $y_i$ as the perpendicular speed of the
quark $v_{i\bot}$, we can write $\mu_{i0}=W_{ir}\gamma_{i\bot}$
with $W_{ir}=\sqrt{p_r^2+m_i^2}$ and
$\gamma_{i\bot}=1/\sqrt{1-v_{i\bot}^2}$, in the usual relativistic flux
tube notations \cite{olss95}. The set of equations~(\ref{seteq1}) can
then be recast into the form
\begin{subequations}
\label{rftset}
\begin{eqnarray}
P_{\bot}&=&0=W_{1r}\gamma_{1\bot}v_{1\bot}-W_{2r}
\gamma_{2\bot}v_{2\bot}+\frac{ar}{v_{1\bot}+v_{2\bot}}\left(\sqrt{1-v^{2
}_{1\bot}}-\sqrt{1-v^{2}_{2\bot}}\right), \label{p0fin} \\
\frac{L}{r}&=&W_{1r}\frac{\gamma_{1\bot}v_{1\bot}^{2}}{v_{1\bot}+v_{2
\bot}}+W_{2r}\frac{\gamma_{2\bot}v_{2\bot}^{2}}{v_{1\bot}+v_{2\bot}}+
\frac{ar}{(v_{1\bot}+v_{2\bot})^{2}}\left(F(v_{1\bot})+F(v_{2\bot
})\right), \label{lorfin} \\
H&=&\gamma_{1\bot}W_{1r}+\gamma_{2\bot}W_{2r}+\frac{ar}{v_{1\bot}+v_{2
\bot}}(\arcsin v_{1\bot}+\arcsin v_{2\bot}),\label{hfin}
\end{eqnarray}
with
\begin{equation}
\label{fy2}
F(v_{i\bot})=\frac{1}{2}\left[\arcsin
v_{i\bot}-v_{i\bot}\sqrt{1-v_{i\bot}^{2}} \right].
\end{equation}
\end{subequations}
These equations are exactly the asymmetric relativistic flux tube
equations of Ref.~\cite{olss95}. Let us note that, in
Ref.~\cite{guba94}, conditions about the extremal values of the
auxiliary fields $\mu_i$ and $\nu$ are given. It seems that they are
obtained assuming that $\zeta$ is independent of these fields. We see
here that the calculation can be completed without this assumption.

The equations obtained are the classical ones. For practical
calculations, they must be quantized. This can be done in three steps
\cite{olss95}:
\begin{enumerate}
\item to replace $L$ by $\sqrt{\ell(\ell+1)}$, where $\ell$ is the
orbital angular momentum quantum number;
\item to replace $p_r$ by the operator
$-\dfrac{1}{r} \dfrac{\partial^2}{\partial r^2} r$;
\item to symmetrize the various non-commuting operators.
\end{enumerate}

The contribution of a potential $V$, simulating effects not taken into
account by the rotating string, can be considered in the starting
Lagrangian~(\ref{nambu}). In this case, the calculation shows that it is
simply added to the
Hamiltonian in Eq.~(\ref{hfin}). Such an interaction must depend on the
module of the relative separation (\ref{cmr}) between quarks. With the
equal time approximation, we have simply $V=V(\vec r\,^2)$.

It is shown here that the rotating string model with the straight-line
ansatz \cite{guba94} reduces
exactly to the relativistic flux tube model \cite{olss95}, when all
auxiliary fields are correctly eliminated. This result generalizes the
work of Ref.~\cite{sema04} to the case of two different quark masses.
Again a clear interpretation can be done for the various characteristic
fields of the rotating string Hamiltonian.

Despite these considerations, the rotating string
equations~(\ref{seteq1}) are nevertheless
interesting to consider, because analytical approximated solutions and
quite precise numerical solutions (using the WKB approximation) can be
obtained for these equations \cite{buis05}. These solutions are good
approximations of the solutions of the genuine relativistic flux tube
equations~(\ref{rftset}).

The authors would like to thank Fabian Brau for helpful discussions. C.
Semay (FNRS Research Associate position) thanks the FNRS for financial
support.


\end{document}